# Model for the structure function constant for index of refraction fluctuations in Rayleigh-Benard turbulence

Robert A. Handler[1,2], Richard J. Watkins[3], Silvia Matt[4], K. P. Judd[5]

1) Department of Mechanical Engineering, George Mason University, Fairfax, Virginia, 22030
2) Center for Simulation and Modeling, George Mason University, Fairfax, Virginia, 22030

3) Micro-Photonics Laboratory, the Holcombe Department of Electrical and Computer Engineering, Center for Optical Materials Science and Engineering Technologies (COMSET), Clemson University, Clemson, South Carolina, 29634, USA

4) Naval Research Laboratory, Stennis Space Center, MS, 39529

5) Naval Research Laboratory, Washington, DC, 20375

**Abstract:** A model for the structure function constant associated with index of refraction fluctuations in Rayleigh-Benard turbulence is developed. The model is based upon the following assumptions: (1) the turbulence is homogeneous and isotropic at or near the mid-plane, (2) the rate of production is in balance with the rate of dissipation, (3) an inertial region exists, and (4) estimates for the rate of dissipation of temperature fluctuations and of turbulent kinetic energy can be made by assuming that the large-scale turbulence is dissipated in one eddy turnover time. From these assumptions, the dependence of the structure function on the geometry, heat flux, and the properties of the fluid is obtained. The model predicts that the normalized structure function constant is independent of the Rayleigh number. To verify the model, numerical simulations of Rayleigh-Benard turbulence were performed using two different approaches: an in-house code based on a pseudo-spectral method, and a finite volume code which employs a model for the smallest scales of the turbulence. The model was found to agree with the results of the simulations, thereby lending support for the assumptions underlying the theory.

# 1. Introduction

Inhomogeneities in the index of refraction field resulting from turbulent temperature fluctuations are a common occurrence in both the atmosphere and hydrosphere. These fluctuations are a contributor, in addition to scattering and absorption, to the degradation of the propagation of light. One such well-known effect attributed to atmospheric turbulence is the twinkling of stars. In this case, the intensity level of the light is observed to vary in an irregular manner (scintillation). The resulting index of refraction fluctuations are frequently referred to as optical turbulence. Light fields propagating through optical turbulence suffer from a spreading effect (beyond diffraction), fluctuations in the light beam position referred to as beam wander, and irradiance variation [1]. Thus, optical turbulence imposes limits on the performance capabilities of imaging systems, sensor networks, directed-energy weapons, and laser-based communications (*Free Space Optics-FSO*). Furthermore, insight to the interactions of light propagation through a turbulent medium using physics-based approaches may inform higher fidelity models for the prediction of optical system performance parameters. This could precipitate the development of advanced real-time adaptive optical systems that continually adjust their performance characteristics to compensate for signal deterioration induced by evolving or degraded environmental conditions.

Field experimentation, both in the atmosphere and the water column, has provided a wealth of information and contributed to various correlations between the environmental state and optical propagation parameters [2-6]. However, experiments are often cost prohibitive, the environmental conditions are typically uncontrolled, and may only provide limited spatial information from a finite number of deployed instrumented monitoring stations. Numerical computations are proving to be a cost effective and powerful approach to simulating multi-physics phenomena whose outcomes may be used to inform and optimize experimental efforts. For this investigation, the Rayleigh-Bénard (RB) turbulent convection problem is the framework in which we study light propagation. Turbulent RB convection serves as a canonical flow model for various environmental areas of research and technological applications [7]. In addition, the model configuration has many attractive qualities for both experimental and simulation approaches such as a simple geometrical setup, well defined boundary conditions, well understood stability and transition characteristics, and rich dynamical behavior [8,9].

In light propagation studies through a turbulent medium, the magnitude of the index of refraction structure constant ($C_n^2$) is often reported since it provides a level of the variation in the refractive index, and hence an estimate of the *strength* of the turbulence. Representative values, for horizontal propagation in the atmosphere, range from $10^{-17} m^{-2/3}$ for weak turbulence to $10^{-13} m^{-2/3}$ for strong turbulence [1]. A predictive model for the behavior of the index of refraction structure constant may be integrated directly into optical design expressions to estimate aperture size limits, optical element diameters, spot size resolution, and other optical system performance metrics [10].

The purpose of the current work is to determine the efficacy of determining the properties of optical turbulence, and specifically $C_n^2$, through the use of three-dimensional numerical simulations of Rayleigh-Benard turbulence, which serves as a convenient model many for environmental flows driven by buoyancy. The simulations were carried out for air using both a pseudo-spectral method, and a finite-volume method which employs a large-eddy-simulation model of the smallest scales of turbulence. The simulations, which were carried out using identically the same fluid properties, computational domain size, and Rayleigh (Ra) numbers, were used to determine $C_n^2$ over an order of magnitude change in Ra. In addition, a model for $C_n^2$ was developed based on the scaling laws of turbulence in the inertial range of turbulent length scales. The model was found to be in good agreement with the simulations, indicating the efficacy in using numerical simulations to obtain important properties of optical turbulence, as well as a means of verifying theoretical estimates of these properties.

## 2. Problem Formulation

The problem of interest, which is referred to as the Rayleigh-Benard (RB) problem [9], is that of a fluid initially at rest in a gravitational field which is driven by a temperature difference between two parallel plates. In such a fluid, temperature differences give rise to density differences, which in turn give rise to buoyancy forces. When buoyancy forces are sufficiently strong to overcome viscous forces and thermal diffusion, fluid motion results. This problem is generally difficult to solve since the fluid cannot be considered incompressible. However, when temperature induced density differences are small compared to a reference density, the *Boussinesq* approximation is often employed [11]. We employ this approximation in this work.

The Boussinesq equations of motion are given by:

$$\frac{D\boldsymbol{V}}{Dt} = -\rho^{-1}\boldsymbol{\nabla}\mathrm{p} + \nu\nabla^2\,\mathbf{V} + \boldsymbol{f_B},  \tag{1}$$

where $D/Dt$ is the material derivative, $p$ is a modified pressure, $\rho$ is a reference density, $\nu$ is the kinematic viscosity, $\boldsymbol{V} = (u, v, w)$ is the fluid velocity in the $x, y, z$ directions respectively, gravitational forces act in the negative $y$ direction, and $x$ and $z$ denote coordinates in the horizontal plane. In addition, $\boldsymbol{f_B}$ is the buoyancy body force per unit mass given by $\boldsymbol{f_B} = g\beta\theta\boldsymbol{\hat{\jmath}}$ where $g$ is the gravitational acceleration, $\beta$ is the coefficient of expansion of the fluid, $\theta = T - T_0$, where $T$ is the temperature and $T_0$ is a reference temperature, and $\boldsymbol{\hat{\jmath}}$ is a unit vector in the positive $y$ direction. Consistent with the Boussinesq approximation, conservation of mass is given by:

$$\boldsymbol{\nabla} \cdot \boldsymbol{V} = 0,  \tag{2}$$

and the temperature field is governed by:

$$\frac{D\theta}{Dt} = \alpha \nabla^2 \theta, \qquad\qquad (3)$$

where $\alpha = {^k}\!/\!_{\rho c_p}$ is the thermal diffusivity, $k$ is the thermal conductivity, and $c_p$ is the heat capacity at constant pressure. The above formulation leads to three non-dimensional numbers: the Nusselt number, $Nu = QL/k\Delta T$, the Rayleigh number, $Ra = \beta g \Delta T L^3 / \nu \alpha$, and the Prandtl number, $Pr = \nu/\alpha$ , where $L$ is the distance between the plates which have fixed temperatures which differ by $\Delta T$ , and the vertically directed average heat flux from the bottom plate is $Q$. From dimensional arguments we must have $Nu = F(Ra, \text{Pr})$, which implies that the non-dimensional heat flux given by the Nusselt number depends only on the Rayleigh and Prandtl numbers.

## 3. Numerical Simulations

### 3.1 Fluid properties, boundary conditions, and initial conditions

Rayleigh-Benard turbulence in air was simulated by two different numerical schemes: (A) Pseudo-spectral methods using an in-house code, and (B) A finite-volume method which employs a large-eddy-simulation (LES) model using the open source computational fluid dynamics package OpenFOAM toolbox. It should be noted that the spectral simulations used in this work do not use turbulence models and will be referred to as direct numerical simulations (DNS).

The same computational domain size, fluid properties, and boundary conditions were used for both spectral and OpenFOAM simulations. On the top and bottom walls of the domain, no-slip conditions were imposed on the velocity field and the walls were kept at constant temperature. Periodic boundary conditions were applied in the $x$ and $z$ directions.

The properties of air were chosen at 293.15 K, and are as follows: $\nu = 1.516 \times 10^{-5} m^2/s$ , $\alpha = 2.074 \times 10^{-5} m^2/s$, $\beta = 3.411 \times 10^{-3} K^{-1}$, and $k = 2.514 \times 10^{-2} W/(m-K)$. The Prandtl number, specific heat, density, and atmospheric pressure were $Pr = 0.7309$, $c_p = 1007 J/(kg-$ $K$), $\rho = 1.204 \ kg/m^3$, and $p_0 = 1013.25$ millibars. The computational domain dimensions were $L_x = 0.5m$, $L_y = L = 0.1m$ and $L_z = 0.5m$ in the $x, y$, and $z$ directions respectively. In the OpenFOAM simulations, the initial velocity was set equal to zero, and the temperature field was set to its ambient value. In the spectral simulations, the initial velocity components were set equal to small random values, and the initial temperature field was chosen to be the linear conduction profile given by $T(y) = \Delta T(\frac{1}{2} - \frac{y}{L}) + T_0$ where $y = 0$ defines the center of the domain. With this initial condition for the temperature field, the Nusselt number at time $t = 0$ is $Nu = 1$ , since the initial heat flux at the bottom wall is $Q = -k \ \partial T / \partial y = k \Delta T / L$. We note that when $Ra > 1708$ [9] this flow will be unstable to infinitely small disturbances, but the Rayleigh numbers in our simulations were about two to three orders of magnitude greater than this, resulting in self-sustaining turbulence.

## 3.2 Description of DNS and LES simulations

For both DNS and LES simulations, five simulations were performed for $\Delta T = 1°C$, $2°C$, $5°C$ ,$10°C$, and $20°C$, which correspond to Rayleigh numbers of $1.063 \times 10^5$ to $2.126 \times 10^6$. Further details concerning the simulations are given in Appendix A in Table 1. In all simulations, statistics were obtained from uncorrelated realizations of the velocity and temperature fields, after the flow reached a statistically steady state.

The spectral simulations, which employ equations (1-3), were performed on the Clemson University Palmetto Cluster using an in-house pseudo-spectral code [12] which employs Fourier modes in the horizontal $(x - z)$ plane and Chebyshev modes in the vertical $(y)$ direction. The grid resolution was $128 \times 65 \times 128$ in the $x, y$, and $z$ directions respectively, and the time step was $\Delta t = 1.5 \times 10^{-4} s$ . In each simulation, the Rayleigh number was fixed by choosing a temperature difference between top and bottom walls. The physical time duration for each simulation was 240 seconds.

The OpenFOAM, finite volume simulations were performed on Centennial, an SGI ICE XA with 1,848 compute nodes, at the Army Research Laboratory, Department of Defense Supercomputing Resource Center (ARL DSRC), one of the supercomputing centers of the DoD High Performance Computing Modernization Program (HPCMP). OpenFOAM is a Navier-Stokes solver based on the finite volume method. For turbulence modeling, the Large Eddy Simulation (LES) approach was used, where the model solves the filtered Navier-Stokes equations. The filter size is dependent on the grid spacing and the sub-grid scales (SGS) are modeled. In our setup, SGS stresses were modeled using the Wall Adapting Local-Eddy viscosity (WALE) model [13]. The grid resolution was $250 \times 250 \times 250$ in the $x, y$, and $z$ directions for the runs with $\Delta T = 1°C, 2°C, 5°C$. For the cases with higher Rayleigh number, ( $i.e.$, $\Delta T = 10°C, 20°C$) the grid resolution was $250 \times 100 \times 250$. The time step for these simulations was $\Delta t = 1.0 \times 10^{-2} s$ for the lower $Ra$ cases and were variable for the higher $Ra$ simulations with a criterion to keep the CFL (Courant-Fredrichs-Lewy) number less than 1 [14], a requirement for numerical stability.

It should be noted that spectral methods exhibit exponential convergence [15] and have been used with success in simulating a wide variety of turbulent flows [16]. This accounts for the significantly larger number of grid nodes (i.e. smaller spacing between grid nodes) required for the OpenFOAM simulations compared to the spectral simulations.

## 4. Simulation Results

A detailed comparison of the velocity and temperature fields obtained from the DNS and LES [17,18] has been performed. The results show quantitative and qualitative agreement. Therefore, to avoid unnecessary repetition, we will primarily exhibit DNS results unless otherwise stated.

Here, in order to give an overall impression of the flow, we present results for the temporal evolution of the Nusselt number, the relation between the Nusselt and Rayleigh numbers, visualizations of instantaneous snapshots of the temperature and index of refraction fields, and the statistics associated with these fields.

## 4.1  Temporal evolution of the Nusselt number

The evolution of the Nusselt number, which is essentially a proxy for the heat flux across the domain, is shown in Figure 1 for all DNS runs. Since the Rayleigh number for all runs was above the linear stability limit, flow instability was induced by seeding the initial velocity field with small random values. This excites a strong response, as can be seen by the rapid increase of the Nusselt number in the first few seconds of each simulation. The flows are seen to fairly quickly establish a steady-state in which the Nusselt number oscillates randomly about a mean value. All statistics from the DNS runs were obtained by averaging over the last 60 seconds of these simulations.

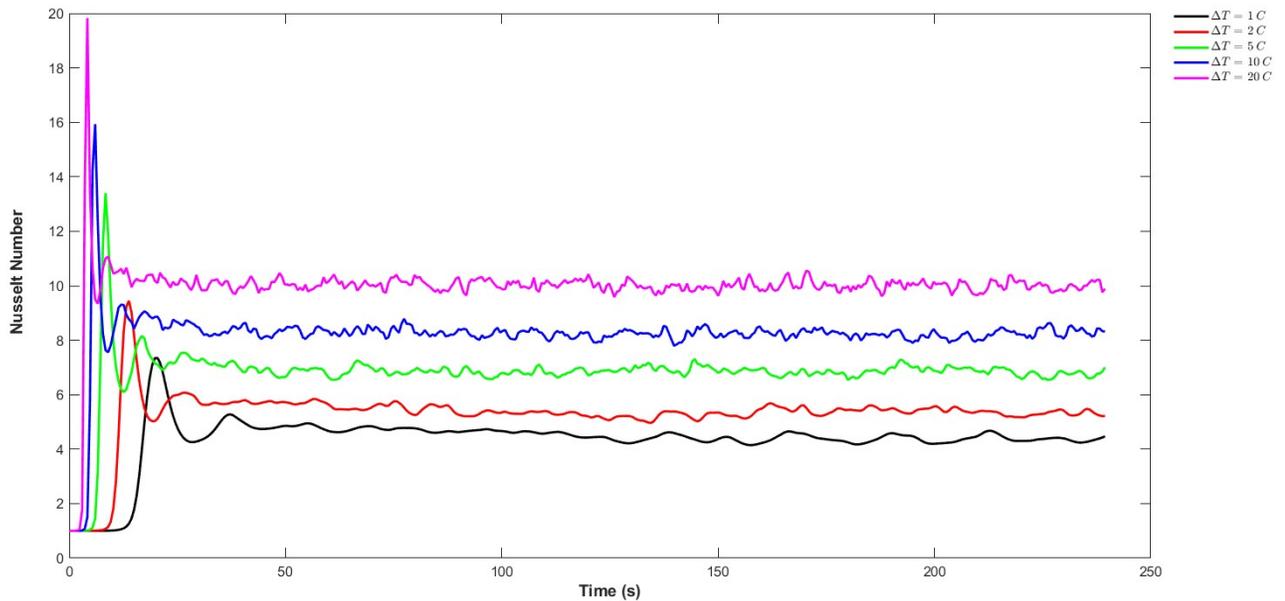

**Figure 1**:  Nusselt number ($Nu = QL/k\Delta T$ ) versus time obtained from the DNS.

## 4.2 Visualization of the temperature field

In Figure 2, a three-dimensional snapshot of the instantaneous temperature field obtained from the last 60 seconds of the DNS for $\Delta T = 20$ °C and $Ra = 2.126 \times 10^6$ is shown. The numerous warm (red) and cold (blue) cusp-like shapes at the bottom and top walls correspond to rising and falling fluid respectively, which was confirmed through numerous visualizations of the velocity field not shown here.

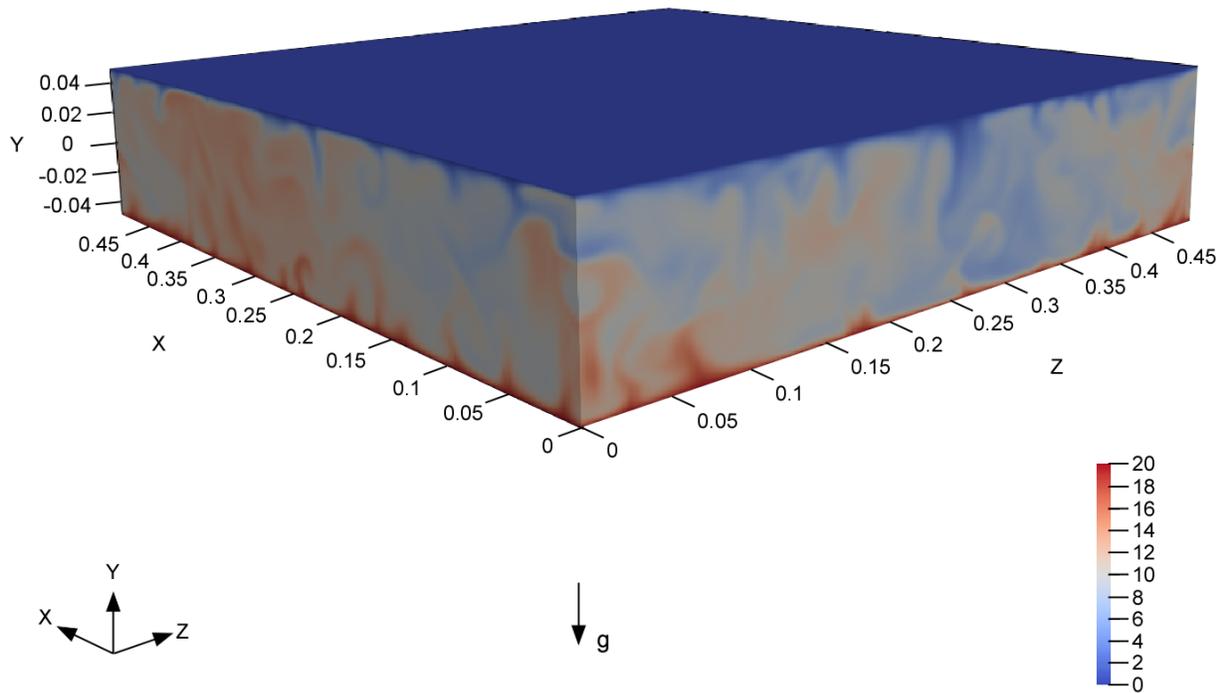

**Figure 2**: Three-dimensional visualization of the temperature field, $\theta = T - T_0$ (°C), obtained from a steady state flow from the DNS for which $\Delta T = 20$ °C and $Ra = 2.126 \times 10^6$ at $t = 180s$. Hot and cold fluid associated with cusp-like regions are rising and falling from bottom and top boundaries respectively. Axes are denoted in meters and the arrow indicates the direction of gravity.

## 4.3 Dependence of the Nusselt number on Rayleigh number:

In Figure 3 the dependence of the Nusselt number on Rayleigh number is shown for both DNS and LES. The results show close agreement over a range of more than one order of magnitude in Rayleigh number, despite the significant differences between the two numerical approaches. In addition, a least squares curve fit using the data from both simulations gives $Nu = 0.186Ra^{0.274}$. This is in reasonable agreement with a comprehensive experiments [19] cited in the classic work of Chandrasekar (1961) ([9]) as representative of laboratory scale RB turbulence.

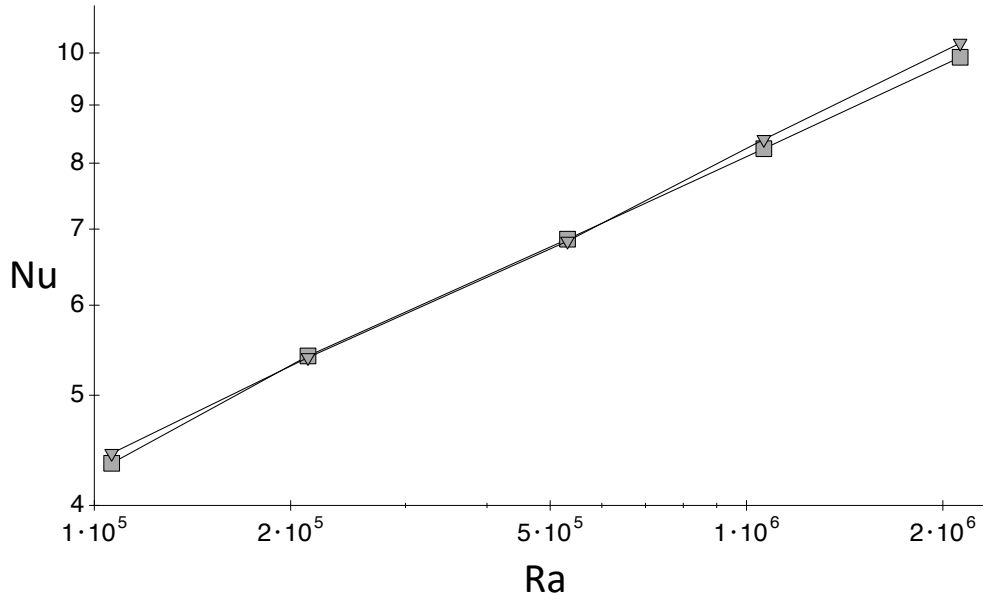

**Figure 3**: Nusselt number vs Rayleigh number. DNS (■) LES (▼). The Nusselt number was obtained by from the heat flux at top and bottom walls of the domain during a time period in which the flow was statistically steady. A least squares curve fit of these results gives $Nu = 0.186Ra^{0.274}$.

## 4.4 Visualization of the index of refraction field

We are concerned with the index of refraction in this work, since its structure function can be used to determine the structure function constant , $C_n^2$. It is straightforward to determine the index of refraction, $n$, in air from the temperature field as follows [1]:

$$n - 1 = -C \frac{p_0}{T_0^2} (T - T_0),$$ (4)

where $C = 79 \times 10^{-6}$, $p_0 = 1013.25$ mbar, and $T_0$ is the reference temperature. The left-hand side represents deviations of the index of refraction from its equilibrium value of 1. In Figure 4 a three-dimensional snapshot of $n_0 = n - 1$ is shown. Here it is evident that $n_0$ is strictly negative since the temperature field is always greater than the reference temperature, which is defined to be the temperature of the top wall. Rising warm plumes (see Figure 2) are seen to be associated with the largest negative index of refraction (e.g. blue cusp-like features in Figure 4).

In Figures 5(a,b) horizontal $(x - z)$ slices of the index of refraction field, obtained from the Figure 4 snapshot, are shown. Figure (5a) shows that the index of refraction field near the top wall exhibits an interesting *spider-web* structure composed of narrow (red) linear features. These features have been confirmed to be associated with falling cold plumes. On the other hand, the index of refraction field at the exact center of the flow shown in Figure 5(b) appears relatively featureless.

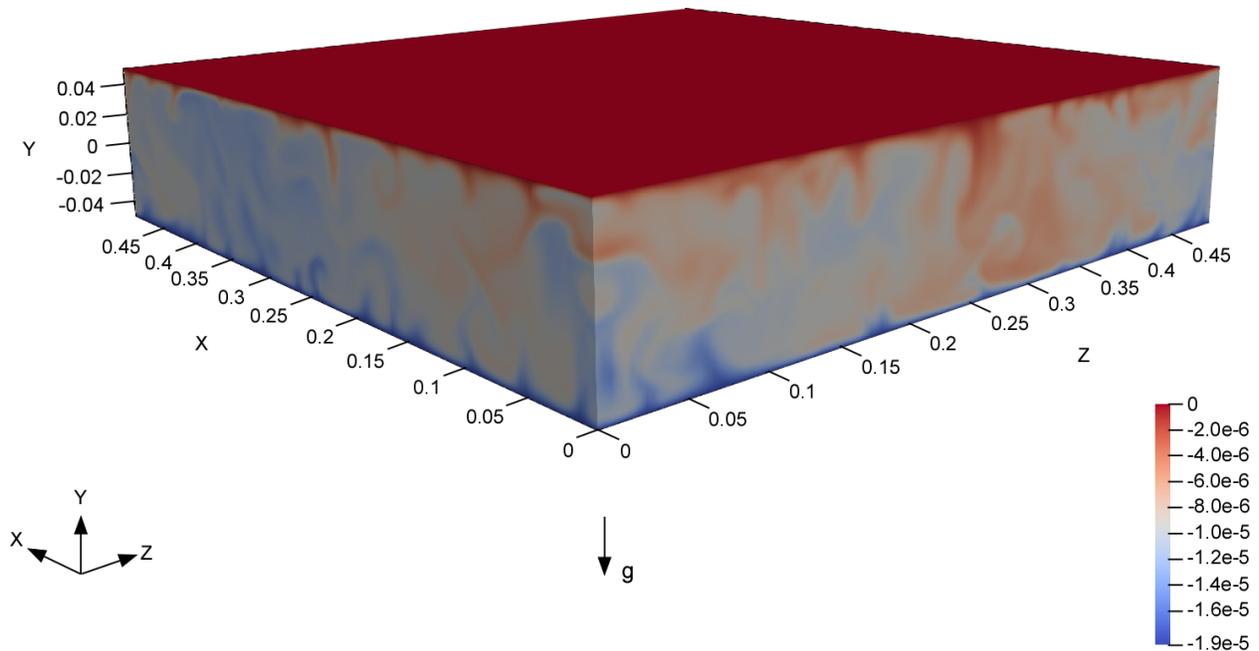

**Figure 4:** Three-dimensional snapshot of the index of refraction field, $n_0 = n - 1$. This image was generated from the thermal field shown in Figure 2.

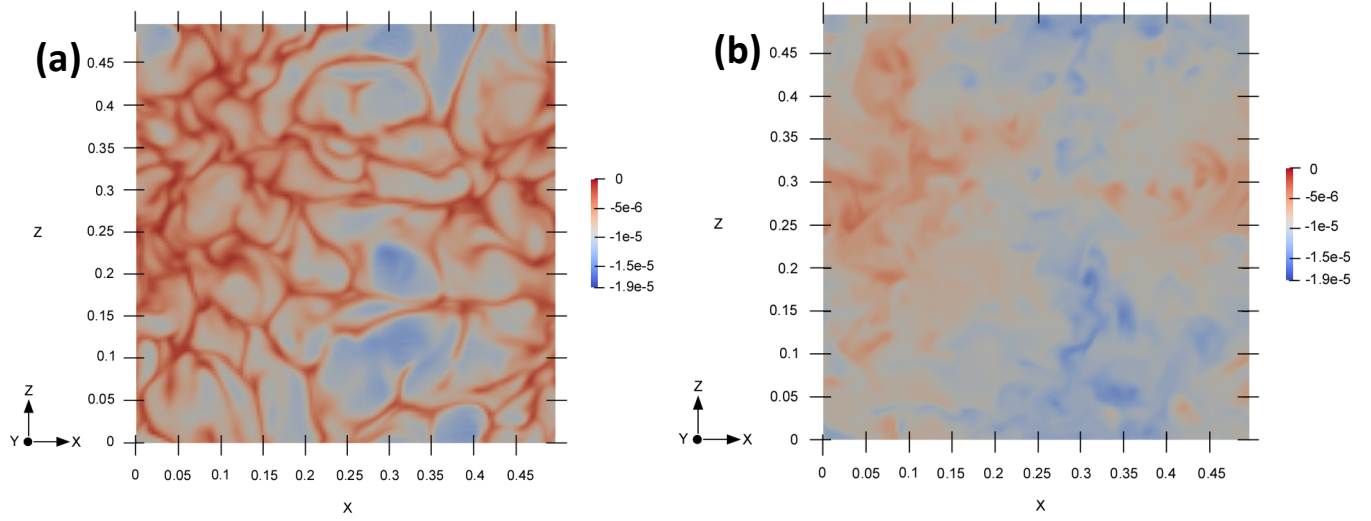

**Figure 5:** **(a)** Visualization of the index of refraction field , $n_0 = n - 1$, in the horizontal $(x - z)$ plane near the top surface $(y = 0.045 \, m)$ obtained from the snapshot shown in Figure 4. Note the "spider-web-like" structure. **(b)** Visualization of the index of refraction field in the horizontal $(x - z)$ plane at the center $(y = 0.0 \, m)$ of the domain.

## 4.5 Statistics of the temperature, velocity, and index of refraction

The statistics of the temperature, velocity, and index of refraction fields are presented in Figures (6-8). Here for any field $\phi$, $\bar{\phi}$ is defined as its average (mean) and is obtained by summing over all realizations of the flow in a given time period and over the horizontal plane. Its root-mean-square (rms) is defined by $\phi_{rms}(y) = [ \overline{(\phi - \bar{\phi})^2} ]^{1/2}$. All statistics presented here were obtained over the time interval $t = 180 - 240 s$ for each simulation. It is expected that the so-called *convective scales* [20] of velocity, $w^*$, and temperature, $\theta^*$, defined by:

$$w^* = (\beta g L q_0)^{1/3} \quad , \tag{5}$$

and

$$\theta^* = q_0/w^*, \tag{6}$$

give reasonable estimates of the velocity and thermal fluctuations for RB turbulence, where $q_0 = Q/(\rho c_p)$, and $Q$ is the heat flux.

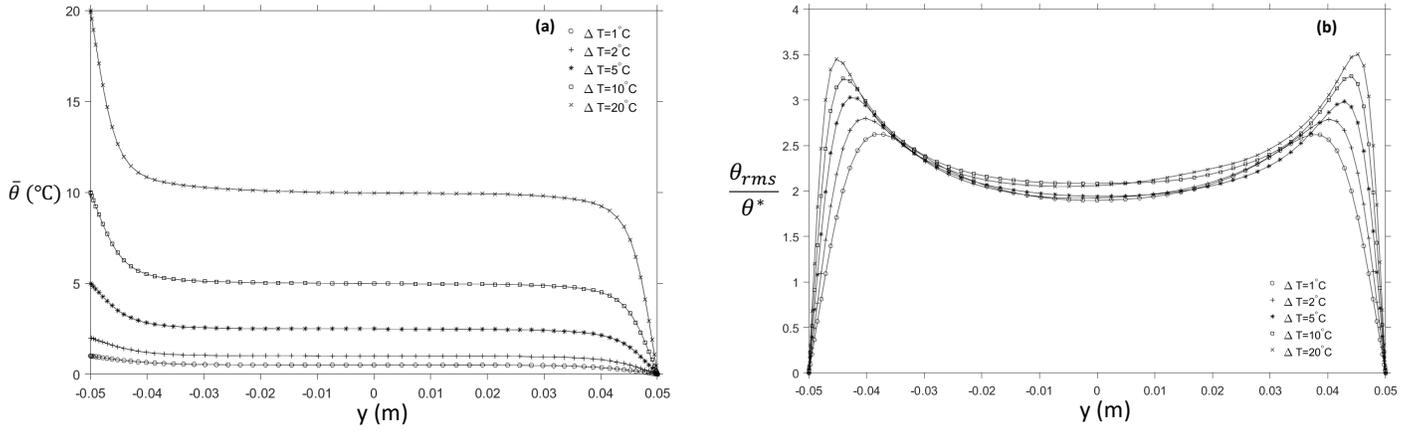

The mean temperature profiles shown in Figure 6(a) indicate the expected deviation from the linear conduction profile used as the initial condition in each simulation as turbulent mixing increases thermal gradients near the walls. As an example, the heat flux for the case $\Delta T = 20°C$ has increased by about one order of magnitude compared to pure thermal conduction as indicated by the fact that the Nusselt is $\mathcal{O}(10)$ in this case, as shown in Figure 3. The rms thermal profiles shown in Figure 6(b) have been scaled using $\theta^*$. This scaling appears to give good data collapse. Further, the thermal fluctuations scaled in this way are all $\mathcal{O}(1)$, which confirms that the thermal convective scale gives reasonable estimates of thermal fluctuation magnitudes. The rms velocity profiles shown in Figure 7 have been scaled using $w^*$. Similar to the thermal fluctuations, all three velocity components are $\mathcal{O}(1)$, and the horizontal $(u, w)$ velocity results show that the flow has no $x - z$ bias as we should expect from the symmetry of the flow. In the center of the flow, although the vertical ($v$) fluctuations are somewhat larger than the horizontal ($u, w$) fluctuations, it is reasonable to assume that the flow can be considered nearly isotropic at its center. In fact, this was also found to be true in recent experiments [21]. Finally, the rms of the index of refraction fluctuations are shown in Figure 8. We note that these profiles, along with those for the velocity and temperature fields, all exhibit maxima near the no-slip walls, where the turbulence is expected to deviate from isotropy.

**Figure 6:** Temperature statistics obtained from the DNS. (a) Mean temperature $\bar{\theta}$ versus vertical distance, $y$, for five temperature differences, $\Delta T$. The bottom wall of the domain is at $y = -0.05m$. (b) Root-mean-square temperature, $\theta_{rms}$, made non-dimensional by the convective thermal scale, $\theta^*$.

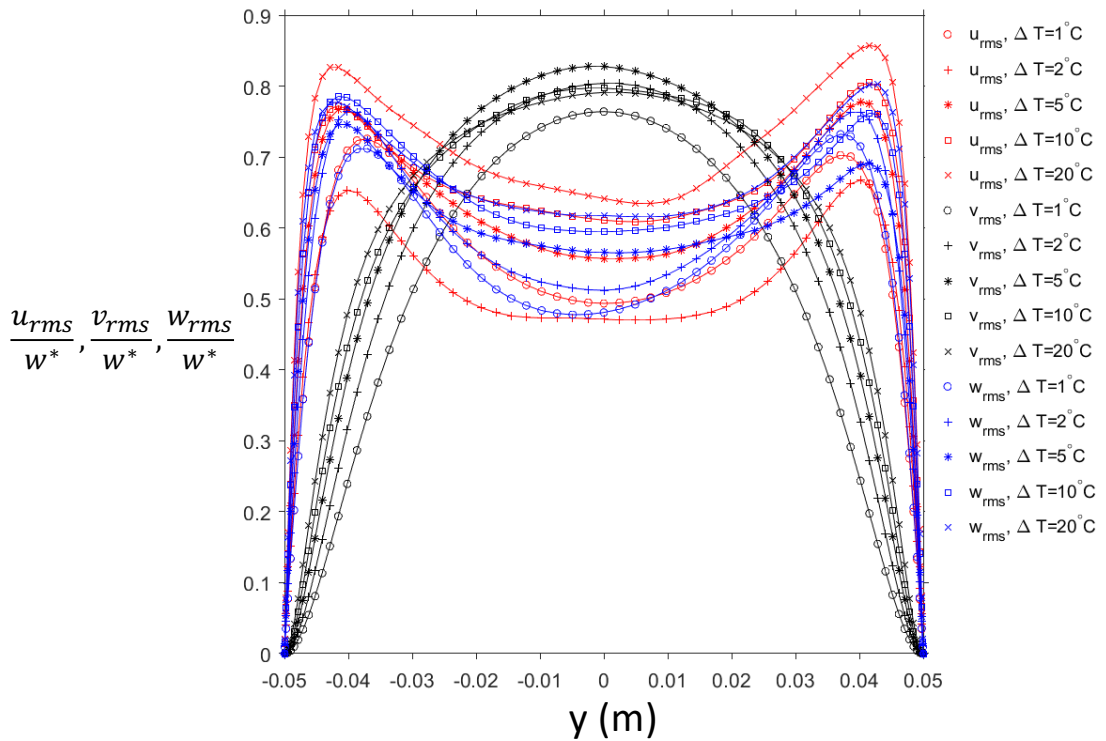

**Figure 7:** Root-mean-square velocities, $u_{rms}$, $v_{rms}$, $w_{rms}$, obtained from the DNS. These have been made non-dimensional by the convective velocity scale, $w^*$.

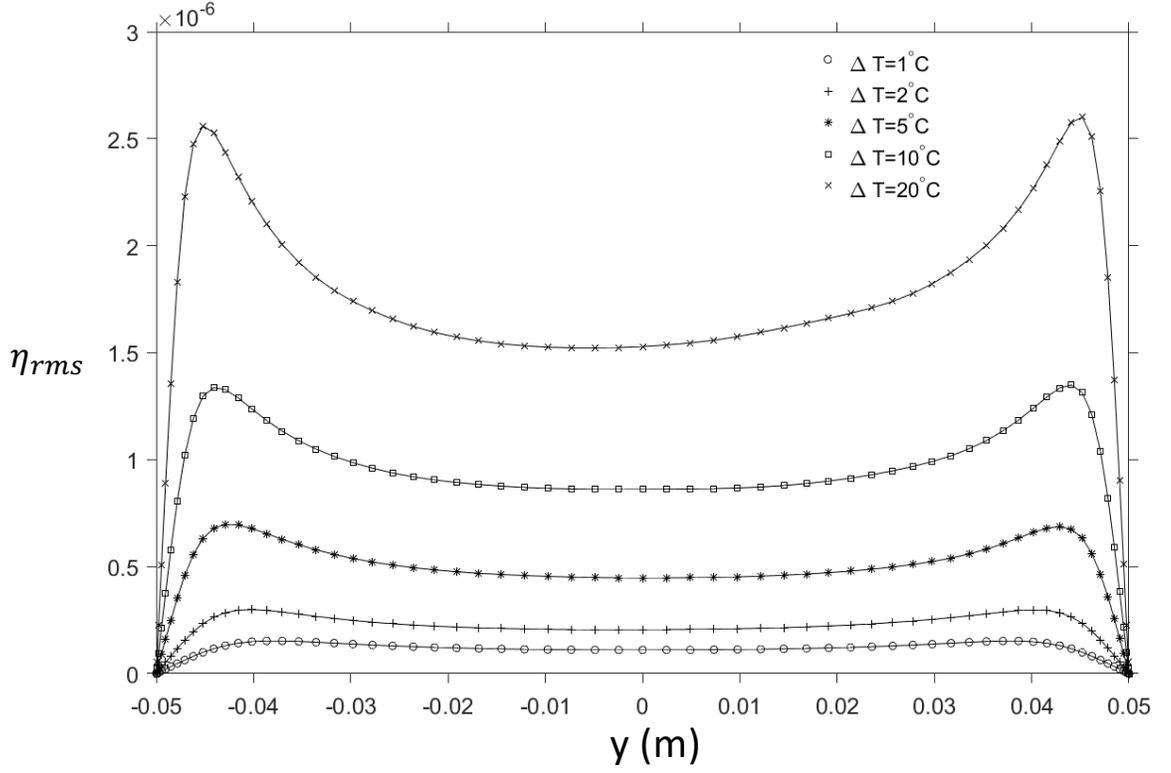

**Figure 8:** Root-mean-square index of refraction , $\eta_{rms}$, obtained from the DNS.

## 5. Model for the structure function constants

In RB turbulence, it is appropriate to define the structure function for temperature fluctuations as a function of $r$, which represents a coordinate in a given horizontal $(x-z)$ plane, and the coordinate, $y$ , which corresponds to that plane as follows:

$$D_T(r,y) = <[(T(0,y) - T(r,y)]^2> \quad , \qquad (7)$$

where the brackets represent averaging over all flow realizations. According to Kolmogorov scaling laws [22] there exists a region such that $l_0 \ll r \ll L_0$ , the so-called inertial region, where $l_0$ and $L_0$ are the inner and outer scales of turbulence for which the structure function is:

$$D_T(r,y) = C_T^2 r^{2/3} \qquad , \qquad (8)$$

where $C_T^2$ is the structure function constant for thermal fluctuations. Similarly, the structure function, $D_n$, for the index of refraction fluctuations is defined by:

$$D_n(r,y) = < [(n(0,y) - n(r,y)]^2 > \qquad . \qquad (9)$$

Since index of refraction fluctuations are assumed to be directly proportional to thermal fluctuations, as a first approximation [1] we must have:

$$D_n(r,y) = K^2 C_T^2 r^{2/3} \qquad , \qquad (10)$$

where $K = -C \frac{p_0}{T_0^2}$, so that an estimate $C_T^2$ will allow us to estimate $C_n^2$ as follows:

$$C_n^2 = K^2 C_T^2 \qquad . \qquad (11)$$

The proposed model for the structure function constants is based on data obtained from the exact center of the simulated RB turbulence where we have shown above that the turbulence is reasonably close to isotropy. Therefore $C_T^2$ can be directly related to the properties of an isotropic, homogeneous turbulence in the inertial range [6] by:

$$C_T^2 \sim \epsilon_T \epsilon^{-1/3} \qquad , \qquad (12)$$

where $\epsilon_T = \alpha < \partial_i T' \partial_i T' >$ is the rate of dissipation of temperature fluctuations, $\epsilon = 2\nu < S_{ij} S_{ij} >$ is the of rate of dissipation of kinetic energy, $S_{ij} = 1/2 \ (\partial_j u_i + \ \partial_i u_j)$ is the rate-of-strain tensor, fluctuations in temperature and velocity are given by $T'$ and $u_i$, and repeated indices imply summation. Furthermore, since the simulated RB turbulence is in a statistically steady state, the production of turbulence equals the rate at which it is being dissipated. In this case, energy is assumed to be supplied to the turbulence at the largest length scales and is dissipated in one large-eddy turnover time, $t_E = L/w^*$, a basic assumption in the Kolmogorov model of turbulence [22]. These considerations lead to the following estimates for the dissipation rates in terms of the convective scales $w^*$ and $\theta^*$ and the $t_E$ :

$$\epsilon_T \sim (\theta^*)^2 w^*/L \qquad , \qquad (13)$$

and

$$\epsilon \sim (w^*)^3 / L \qquad . \qquad (14)$$

The final estimate for $C_T^2$ can be determined by substituting equations (13) and (14) into equation (12) which gives:

$$C_T^2 \sim \left( \frac{Q}{\rho c_p L} \right)^{4/3} (\beta g)^{-2/3} \qquad . \qquad (15)$$

This result implies that

$$\overline{C_T^2} \; = \; C_T^2 \Big/ \left[ \left( \frac{q_0}{L} \right)^{4/3} (\beta g)^{-2/3} \right] \equiv \gamma \qquad , \qquad (16)$$

where $\gamma$ is a dimensionless constant which does not depend on Rayleigh or Prandtl numbers. Naturally, since the Rayleigh number depends on the Nusselt and Prandtl numbers, $\gamma$ should also be independent of the Nusselt number as well. The extent to which this model describes the nature of the structure function constant will be determined by comparing it to simulations.

## 6. Comparison of the model for $C_T^2$ and $C_n^2$ with simulations

For both the DNS and LES simulations, the structure function $D_T$ was determined using equation (7) and data from the center plane of the computational domain. Then using equation (8), the best fit of the structure function to the two-thirds law was used to determine the structure function constant $C_T^2$. In all cases, the goodness of fit parameter, $R^2$, was greater than 0.95 indicating that the two-thirds law was well satisfied.

In Figure 9 (a,b) the dependence of $C_T^2$ on Nusselt and Rayleigh numbers are shown for both DNS and LES simulations. These results are compared with the theory given by equation (15). It is evident that the simulation results and the theory show the same dependence on Nusselt and Rayleigh numbers since both simulation and theory give nearly straight lines with virtually the same slopes as shown using log-log axes. Furthermore, when $C_T^2$ is made non-dimensional as described in equation (16), it is clear from Figure 10 that $\overline{C_T^2}$ is essentially independent of Nusselt and Rayleigh numbers, as predicted by the theory.

Naturally, we expect some variation in $\gamma$ which is evident in the results. Using the DNS and LES data, the average and root-mean-square values for $\gamma$ are found to be $\bar{\gamma} = 6.171$ and $\gamma_{rms} = 0.4721$ respectively. The final result for the normalized thermal structure function constant for RB turbulence is then:

$$\overline{C_T^2} = 6.171 \pm 0.4721 \quad . \tag{17}$$

The dependence of the index of refraction constant $C_n^2$ on heat flux is shown in Figure 10. Here the theory given by $C_T^2 = \bar{\gamma} K^2 \left(\frac{Q}{\rho c_p L}\right)^{4/3} (\beta g)^{-2/3}$ is compared to the simulations. Excellent agreement is obtained, confirming that the model given by equation (15) gives an accurate estimate of the index of refraction structure function constant. We note that although we have shown that $\overline{C_n^2}$ is independent of the Nusselt and Rayleigh numbers, we have not shown that it is independent of the Prandtl number. This dependence, if any, could not be determined based on our current simulations which were performed at constant Prandtl number.

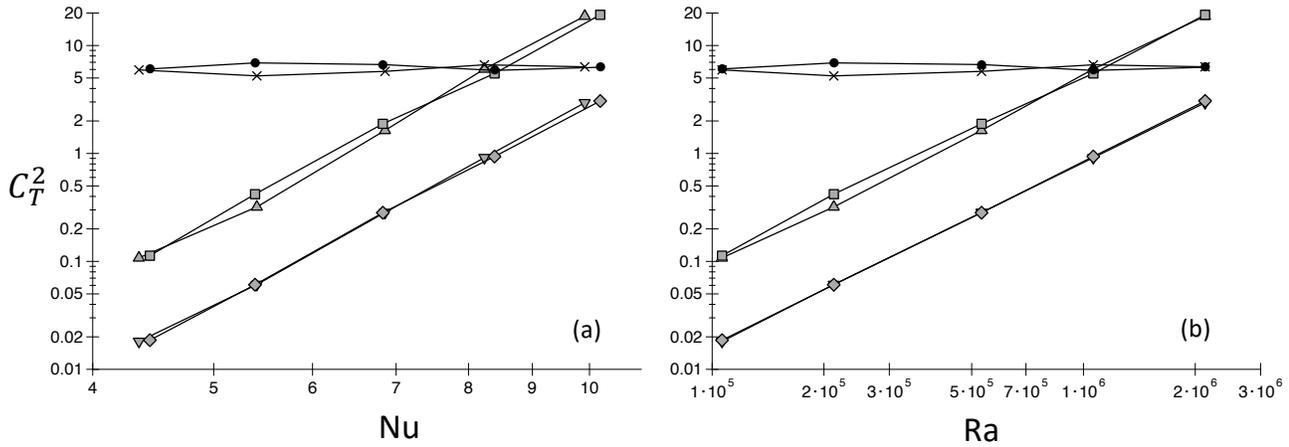

**Figure 9:** Dependence of thermal structure function constant, $C_T^2$, on the Nusselt number (a) and Rayleigh number (b). Theory given by $C_T^2 = \left(\frac{Q}{\rho c_p L}\right)^{4/3} (\beta g)^{-2/3}$ using the $Q$ from DNS ( ▼ ) and LES ( ◆ ). $C_T^2$ from DNS ( ▲ ) and LES ( ■ ). Non-dimensional thermal structure function constant $\overline{C_T^2}$ for DNS (×) and LES ( ● ). Units for $C_T^2$ are $°C^2 m^{-2/3}$.

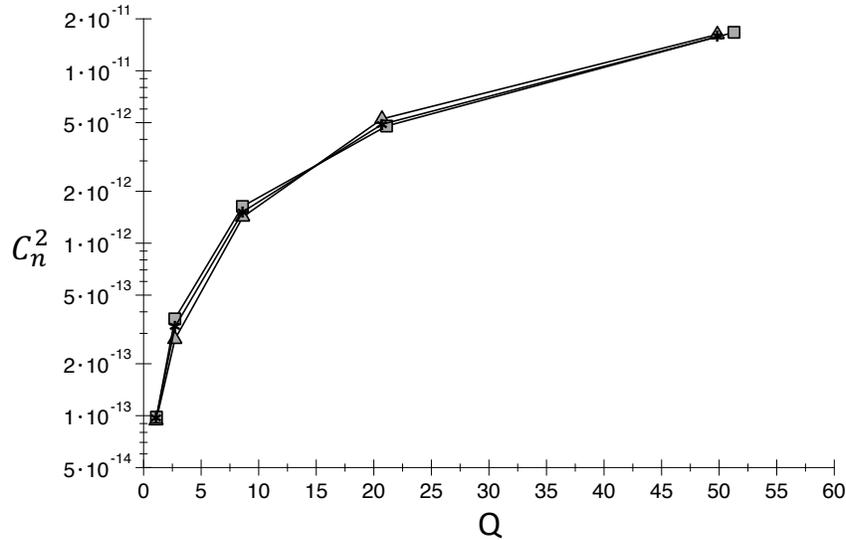

**Figure 10:** Dependence of the index of refraction structure function constant, $C_n^2 = K^2 C_T^2$ where $K = -C\frac{p_0}{T_0^2}$, on heat flux, $Q$ (W/m²). Simulation results: DNS ( ▲ ) , LES ( ■ ). Theory ( ∗ ) given by $C_T^2 = \bar{\gamma} K^2 \left(\frac{Q}{\rho c_p L}\right)^{4/3} (\beta g)^{-2/3}$ . Units for $C_n^2$ are $m^{-2/3}$.

## 7. Summary and Discussion

The objective of this effort was to determine the efficacy of determining the properties of optical turbulence through the use of three-dimensional numerical simulations. The classic and well-studied case of Rayleigh-Benard turbulence was chosen for the simulations since this form of turbulence has many of the features we expect to find in natural flows in atmospheres and oceans, and is therefore of particular interest to the optics and fluid mechanics communities. The simulations were used to study the nature of the structure function constant, $C_n^2$. This statistic is a measure of the intensity of optical turbulence, and therefore determines the effects of thermal fluctuations on the propagation of light through random media.

The simulations were carried out for air using two different numerical schemes: (A) Pseudo-spectral methods using an in-house code (DNS), and (B) A finite-volume method which employs a large-eddy-simulation (LES) model using the OpenFOAM toolbox. The DNS and LES were carried out in identically the same computational domain (aspect ratio of $5:1$), with identically the same fluid properties, and for the same temperature changes across the domain ($\Delta T = 1°C, \ 2°C, \ 5°C, \ 20°C$). This resulted in a Rayleigh number variation of $1.063 \times 10^5$ to $2.126 \times 10^6$. The statistics (e.g. velocity, temperature) obtained from the DNS and LES simulations were compared in previous work, and were shown to be in very close agreement. It is important to emphasize that such close agreement was obtained despite the fact that the DNS and LES use different representations for the small scales of turbulence, and were performed on entirely different computational platforms. In fact, the DNS uses no turbulence models, and therefore uses no special representation of the smallest scales of turbulence.

Three-dimensional visualizations of the temperature and index of refraction fields show a dendritic or *spider-web* structure near the walls, and a more featureless structure at the center of the domain.

Root-mean-square (rms) statistics of the velocity field reveal an isotropic structure at the domain center, and are shown to collapse reasonably well by using the convective scale $w^*$. Similarly, the rms statistics of the temperature field are shown to collapse using the scale $\theta^*$.

A model for the structure function constant for RB turbulence is proposed based on the idea that the turbulence is nearly isotropic at its center, which allows the standard Kolmogorov theory of turbulence to be invoked. Based on this idea, the structure function constant for the thermal fluctuations, $C_T^2$, can be represented in the inertial range by $C_T^2 \sim \epsilon_T \epsilon^{-1/3}$, where $\epsilon_T$ and $\epsilon$ are the dissipation rates for the temperature and kinetic energy respectively. The dissipation rates as well as the large eddy turnover time can then be expressed in terms of the convective scales to give $\overline{C_T^2} = C_T^2 / \left[ \left( \frac{q_0}{L} \right)^{4/3} (\beta g)^{-2/3} \right] \equiv \gamma$, where $\gamma$ is a constant, independent of Rayleigh or Prandtl numbers. Excellent agreement was found between this theoretical result and the simulations. The comparison gives the final result of this work: $\overline{C_T^2} = 6.171 \pm 0.4721$. From this, $C_n^2$ can be determined since it is directly proportional to $C_T^2$. We emphasize that the theory predicts that $\gamma$ should be independent of Rayleigh number, which is confirmed by the simulations for a one order of magnitude change in the Rayleigh number.

In conclusion, this work shows the efficacy in using numerical simulations to obtain important properties of optical turbulence, as well as a means of verifying theoretical estimates of these properties. We also recognize the importance of experimentally verifying the results obtained in this work. In particular, the theory predicts that the structure function constant is related to the heat flux, $Q$, as follows: $C_T^2 \sim Q^{4/3}$. An experimental effort with this goal in mind is now underway in the Clemson University VTG (Variable Turbulence Generator) facility [23,24].

## Acknowledgements:

R.A.H. and R.J.W. acknowledge funding provided by ONR MURI N00014-20-1-2558 and the generous allotment of computer time on the Clemson University Palmetto cluster. This work was also supported in part by an NRL base program. Numerical simulations with OpenFOAM were performed on resources provided by the DoD Supercomputing Resource Center (ARL DSRC).

The data that support the findings of this study are available from the corresponding author upon reasonable request.

# Appendix A: Numerical simulation parameters

**Table 1:** List of simulation parameters for spectral and finite volume simulations.

| Numerical Method | Temperature Difference ($\Delta T$°C) | Rayleigh Number ($Ra$) | Grid Size ($x, y, z$) | Time Step (secs.) | Total Simulation Time (secs.) |
|---|---|---|---|---|---|
| Pseudo-Spectral (DNS) | 1.0 | $1.063 \times 10^5$ | $128 \times 65 \times 128$ | $1.5 \times 10^{-4}$ | 240 |
| | 2.0 | $2.126 \times 10^5$ | $128 \times 65 \times 128$ | $1.5 \times 10^{-4}$ | 240 |
| | 5.0 | $5.315 \times 10^5$ | $128 \times 65 \times 128$ | $1.5 \times 10^{-4}$ | 240 |
| | 10.0 | $1.063 \times 10^6$ | $128 \times 65 \times 128$ | $1.5 \times 10^{-4}$ | 240 |
| | 20.0 | $2.126 \times 10^6$ | $128 \times 65 \times 128$ | $1.5 \times 10^{-4}$ | 240 |
| Finite Volume (LES) | 1.0 | $1.063 \times 10^5$ | $250 \times 250 \times 250$ | $1.0 \times 10^{-2}$ | 75 |
| | 2.0 | $2.126 \times 10^5$ | $250 \times 250 \times 250$ | $1.0 \times 10^{-2}$ | 75 |
| | 5.0 | $5.315 \times 10^5$ | $250 \times 250 \times 250$ | $1.0 \times 10^{-2}$ | 75 |
| | 10.0 | $1.063 \times 10^6$ | $250 \times 100 \times 250$ | variable | 35 |
| | 20.0 | $2.126 \times 10^6$ | $250 \times 100 \times 250$ | variable | 35 |